\newcommand{\Lya}{Ly$\alpha$}
\newcommand{\Lyb}{Ly$\beta$}
\newcommand{\cq}{C$\:${\small IV}}

\newcommand{\cdue}{C$\:${\small II}}
\newcommand{\cuno}{C$\:${\small I}}
\newcommand{\siq}{Si$\:${\small IV}}
\newcommand{\sitre}{Si$\:${\small III}}
\newcommand{\sidue}{Si$\:${\small II}}
\newcommand{\mgd}{Mg$\:${\small II}}
\newcommand{\fed}{Fe$\:${\small II}}
\newcommand{\nc}{N$\:${\small V}}

\newcommand{\huno}{H$\:${\small I}}
\newcommand{\osei}{O$\:${\small VI}}
\newcommand{\aldue}{Al$\:${\small II}}
\newcommand{\altre}{Al$\:${\small III}}
\newcommand{\cadue}{Ca$\:${\small II}}
\newcommand{\nauno}{Na$\:${\small I}}
\newcommand{\cm}{cm$^{-2}$}
\newcommand{\kms}{km s$^{-1}$}
\newcommand{\lsim}{\raisebox{-5pt}{$\;\stackrel{\textstyle <}{\sim}\;$}}

\documentstyle[11pt,aaspp4]{article}

\slugcomment{submitted to the Astronomical Journal \today}

\lefthead{Cristiani and D'Odorico}
\righthead{High-Resolution Spectroscopy of the
HDF-S QSO J2233-606}

\begin{document}

\title{High-Resolution Spectroscopy from 3050 to 10000 \AA\ of the HDF-S
QSO J2233-606 with UVES at the ESO VLT \footnote{Based on material
collected with the ESO VLT telescope}}

\author{S. Cristiani \altaffilmark{2}}
\affil{ ST European Coordinating Facility,
European Southern Observatory, D-85748 Garching bei M\"unchen, Germany}

\and

\author{V. D'Odorico}
\affil{Institut d'Astrophysique de Paris, 98bis Bd Arago, F-75014 Paris
} 


\altaffiltext{2}{On leave from the Astronomy Department, University of Padua.}


\begin{abstract}
We report on high-resolution observations ($\Re \simeq 45000$) of the
Hubble Deep Field South QSO J2233-606 obtained with the VLT UV-Visual
Echelle Spectrograph (UVES). We present spectral data for the
wavelength region $3050 < \lambda < 10000$ \AA. The $S/N$ ratio of the
final spectrum is about 50 per resolution element at 4000 \AA, 90 at
5000 \AA, 80 at 6000 \AA, 40 at 8000 \AA.
Redshifts, column densities and Doppler widths of the absorption
features have been determined with Voigt-profile fitting.
A total of 621 lines have been measured. In particular
270 \Lya\ lines, 41 \Lyb\ and 24 systems containing metal lines have
been identified.  Together with other data in the literature, the
present spectrum confirms that the evolution of the number density of \Lya\
lines with $\log N($\huno$) > 14$ has an upturn at $z \sim 1.4-1.6$.

\end{abstract}




%
\section{Introduction}

Starting on September 28, 1998 and for two weeks, the Hubble Space
Telescope aimed at the same narrow slice of sky in the constellation
Tucana.  The observing strategy of the Hubble Deep Field South
(HDF--S, \cite{williams00}) differs from its northern analogous in
several respects.  
The Space Telescope Imaging
Spectrograph field was centered on a relatively bright ($B \simeq 17.5$) 
quasar at intermediate redshift (J2233-606, $z_{\rm em} = 2.238$).
The installation of STIS and NICMOS on HST in 1997
February has enabled parallel observations with three cameras. In this
way the HDF--S dataset includes deep WFPC2 imaging
(\cite{casertano00}), UV-Visible imaging 
(\cite{gardner00}) and spectroscopy (\cite{savaglio99}), 
deep near-infrared imaging (\cite{fruchter00}),
and wider-area flanking field observations (\cite{lucas00}).
Ground-based observations have been carried out at ESO with the VLT
and NTT telescopes in the framework of the ESO HDF-South Project
({\tt http://www.eso.org/paranal/sv/svhdfs.html}).
The simultaneous availability of
deep imaging and a large spectroscopic coverage at medium-high
resolution makes the HDF--S a unique field to study the relationship
between galaxies and absorbers, the quasar environment, the abundance
pattern of metal absorption systems.

Spectroscopic observations of J2233-606 have already been reported by
Savaglio (1998, SAV98), Sealey et al. (1998), Outram et al. (1999,
OUT99), and Savaglio et al. (1999).
The relationship between the low redshift ($z \le 1$) galaxies lying
within 1' of the QSO and the absorption systems is described by Tresse
et al. (1999), who also found another QSO in the
field at an angular separation of 44.5'' from the HDF-S QSO, with 
$z_{\rm em} = 1.336$.

In this paper we present high-resolution spectroscopy ($\Re \simeq
45000$) carried out with the VLT UV-Visual Echelle Spectrograph,
UVES (\cite{dekker00}). 
These new data are unique in terms of $S/N$ and spectral range
(3050-10000 \AA) and ideally match the HST STIS observations obtained
in the interval 2275-3118 \AA\ with a FWHM resolution of 10 \kms\ 
(\cite{savaglio99}). 
In Sect.~2 the observations and data reduction 
are described. Sect.~3 deals with the process of identification of the 
absorption lines and their fit through $\chi^2$ minimization of Voigt
profiles. In Sect.~4 the general properties of the \Lya\ forest are
discussed, while Sect.~5 describes the 24 identified metal systems.

\section{The Observations}

The UVES observation of J2233-606 were carried out during the
commissioning of the instrument in October 1999.
Details are given in Table~\ref{tab:obs}.
\placetable{tab:obs}
Two set-ups were used: the first (dichroic 1) covering simultaneously
the 3050-3875 \AA\ and 4795-6815 \AA\ ranges in the blue and red arm,
respectively; the second (dichroic 2) covering simultaneously
the 3760-5004 \AA\ and  6725-10000 \AA\ ranges in the blue and red arm,
respectively.

The data have been reduced in the new UVES context available in the
99NOV edition of MIDAS, the ESO data reduction system.  For the
wavelength calibration Thorium lamp spectra have been used.
Wavelengths have then been corrected to vacuum heliocentric values.

The final combined spectrum  has been rebinned to a constant pixel
size of 0.05 \AA\ and covers the wavelength range $\lambda\lambda\,
3050-10000$ \AA. 

The resolution, as measured from the Thorium lines of the
calibration spectra, extracted and treated in the same way 
as the QSOs spectra, is $\Re \simeq 45000$.
The $S/N$ ratio of the final spectrum is about 50 per resolution
element at 4000 \AA, 90 at 5000 \AA, 80 at 6000 \AA, 40 at
8000 \AA.


The continuum level was established selecting regions apparently free
of absorptions and then fitting a cubic spline with a large smoothness
parameter.
The normalized spectra were then spliced together to form a total 
spectrum.

\begin{figure}
\plotone{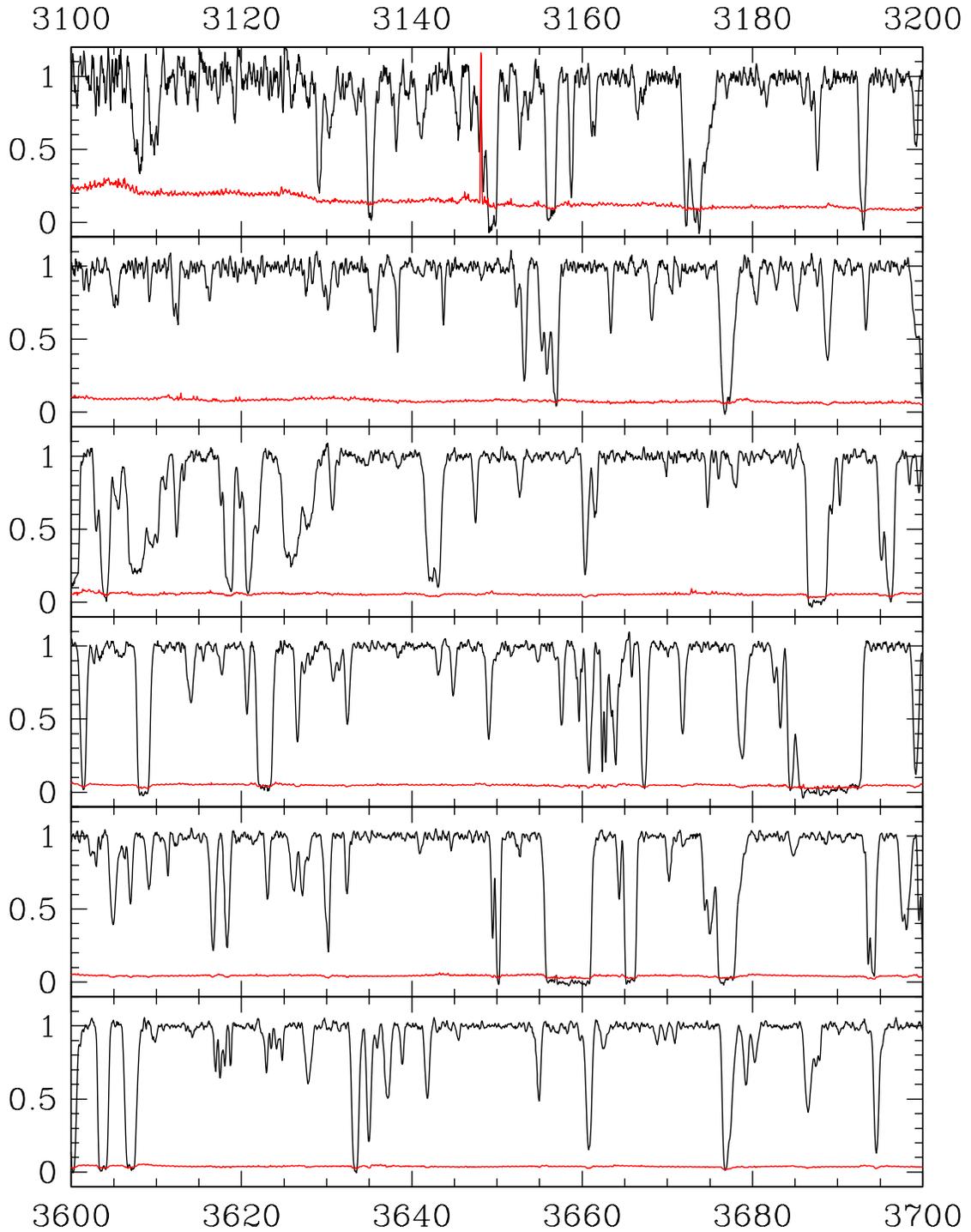}
\caption{The normalized spectrum of J2233-606, plotted against vacuum
heliocentric wavelength (\AA). Each panel contains a strip of 100 \AA\ 
with the wavelength increasing from the top to the bottom. The $1
\sigma$ error level is also shown.
\label{spec1}}
\end{figure}

\begin{figure}
\plotone{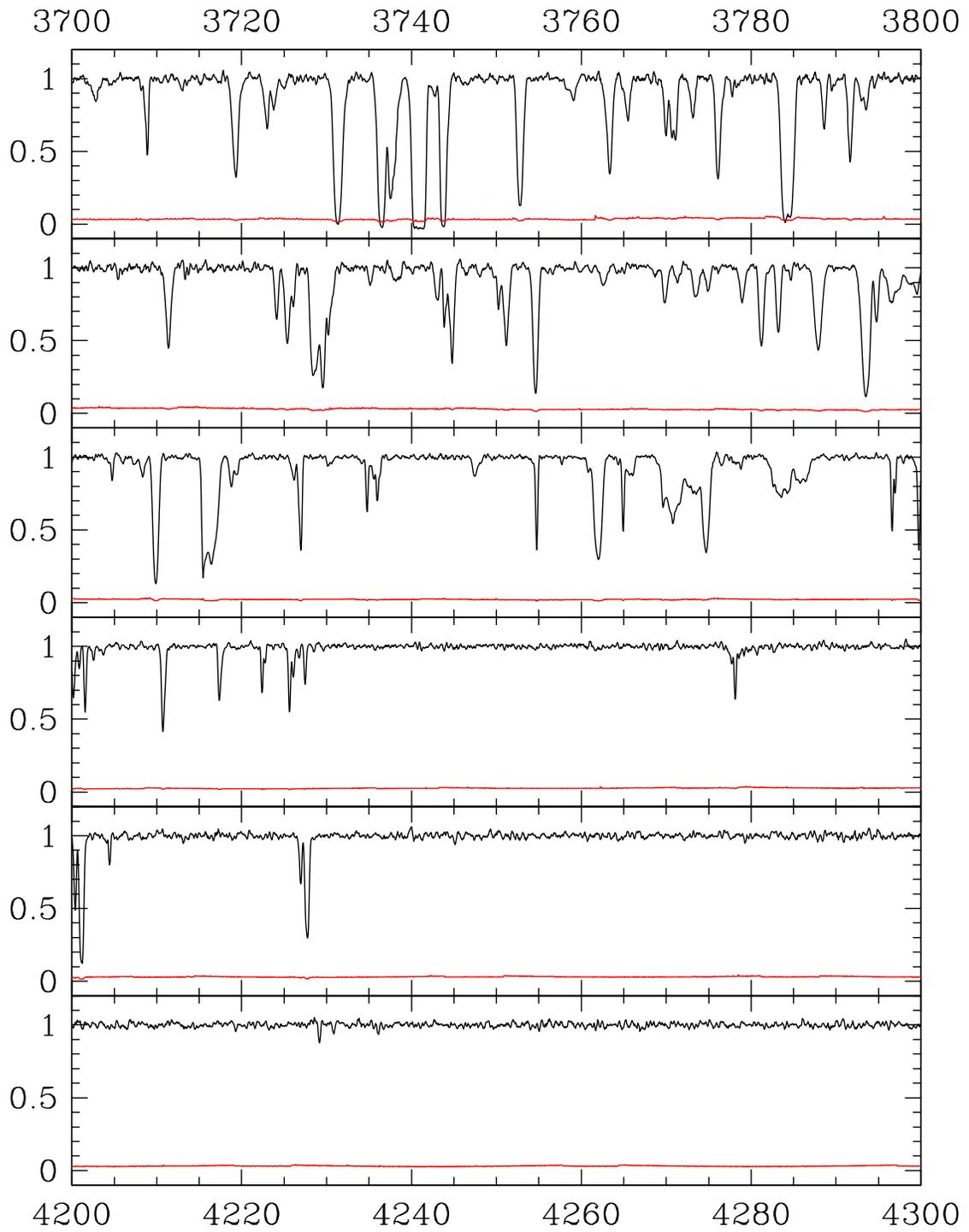}
\caption{The normalized spectrum of J2233-606, plotted against vacuum
heliocentric wavelength (\AA). Details as in Fig.~\ref{spec1}.
\label{spec2}}
\end{figure}

A section of the normalized spectrum from 3100 to 4300 \AA\ 
is shown in Fig.~1-2. 
The full spectrum is available in the form of FITS tables at the URL
{\tt http://www.stecf.org/hstprogrammes/J22/J22.html}.
\placefigure{spec1}
\placefigure{spec2}

\section{Line identification and profile fitting}

The MIDAS package FITLYMAN (Fontana \& Ballester 1995) has been used
to measure the parameters of the absorption lines. Line fitting  through
$\chi^2$ minimization of Voigt profiles has been carried out in order
to determine the redshifts, column densities and Doppler widths
of the identified features.
When possible, we based our fit on the components found for multiple ions
at the same redshift.
The absorption-line parameter fits are presented in Table~\ref{tab:lines}.
Heavy element systems are identified, in general, by the presence of
the \cq\ and/or \mgd\ doublets. The spectral separation and
intensity ratio between the lines of the doublet allow a firm
identification. 
Subsequently, the strongest ions commonly observed in quasar spectra are
searched for at the same redshift. 
The atomic parameters for the absorption lines observed in
QSO spectra have been taken from Morton et al. (1991) and Verner,
Barthel \& Tytler (1994). 
All the lines not identified as metals in the \Lya\ forest are fitted
as \huno\ \Lya\ and \Lyb. 

If a strong ion is not detected in the spectrum, we report only an
upper limit to the column density. The estimate is based on the fact
that there is a linear relation between rest equivalent width and column
density for weak lines (because they lie on the linear part of the
curve of growth). First, the equivalent width limit for that region is
computed and the corresponding column density is given by:

\begin{equation} 
N_{\rm lim}({\rm cm}^{-2}) \simeq 1.13 \times 10^{20}\ w_{\rm lim}(\mbox{\AA})
/ (\lambda_{\rm rest}^2 (\mbox{\AA}) \times f_{\rm osc}).
\end{equation} 

\section{The \Lya\ Forest}

\begin{figure}
\plotone{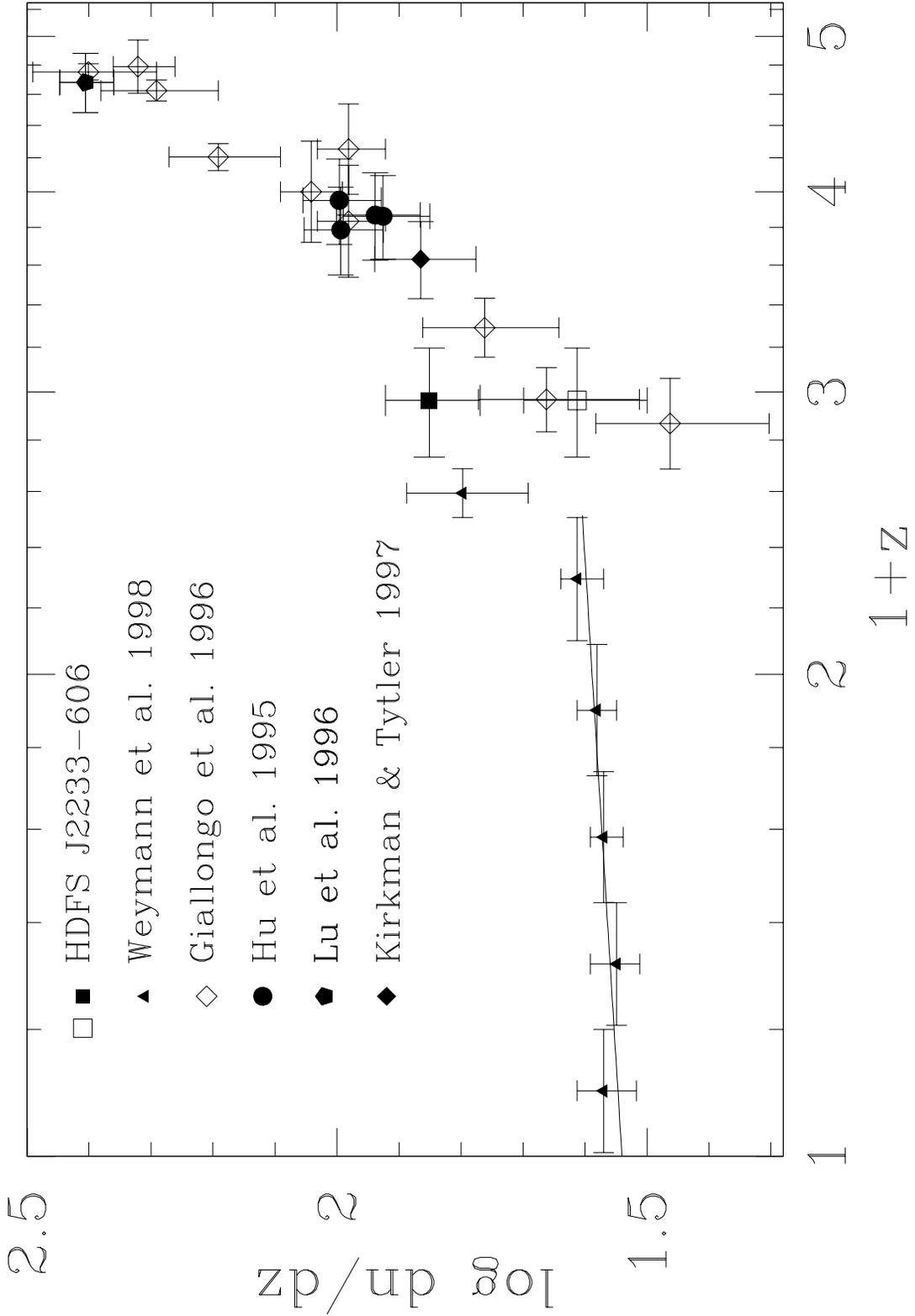}
\caption{Number density evolution of the \Lya\ clouds with $\log
N($\huno$) > 14$ from $z=0$ to $z=4$. Filled symbols are for samples that
include metal systems, open symbols do not. The low-redshift line is
the fit $dn / dz \propto (1+z)^{0.16}$ derived from a sample pf \Lya\
absorbers with equivalent width $EW\ge 0.24$ \AA\ (Weymann et
al. 1998) which correspond to  $\log N($\huno$) > 14$ for $b \sim 26$\kms .
\label{DnDz}}
\end{figure}

A total of 186 \Lya\ lines are observed in the interval
$\lambda\lambda\, 3321.3 - 3885$ \AA, i.e. between the onset 
of the \Lyb\ forest and the range
affected by the proximity effect.
The HI column density distribution for $13 < \log N($\huno$) < 17.5$ 
is consistent with a power-law distribution with a slope
$-1.41\pm0.05$, slightly flatter but not significantly different
from the canonical
value observed at higher redshift (\cite{giallo96}).
The Doppler parameter distribution peaks between 20-25 \kms, with a
lower cutoff at 15 \kms  for lines with  $\log N($\huno$) > 13 $.
\placefigure{DnDz}
Fig.3 shows the number density of \Lya\ lines with $\log N($\huno$) > 14$
observed in the spectrum of J2233-606.  At a medium redshift $<z> =
1.96 \pm 0.23$ we observe 33 such lines, 14 of which are associated
to metal systems (see below).  The corresponding $dn/dz$ is $71 \pm 12
$ including metal systems and $41 \pm 9$ excluding metal systems.
The previous estimate of the $dn/dz$ in J2233-606 by Savaglio et
al. (1999) is in good agreement with the present result.
The comparison with data in the literature from other sight-lines, 
provides a consistent picture of the evolution of the density of \Lya\ 
absorbers with redshift.  
In particular, the measurement derived from
the spectrum of J2233-606 is remarkably close to the HST point at $<z>
= 1.6$ by Weymann et al. (1998), which is based on 19 lines
observed in the spectrum of the QSO UM 18.  Fig.~3 suggests that a
sharp upturn in the density of \Lya\ lines with $\log N($\huno$) >14$
takes place at a redshift $z=1.4-1.6$.

The density of \Lya\ lines in J2233-606 not associated with metal
systems (see next Section) is also shown in Fig.~3.  The agreement
with the corresponding data of Giallongo et al. (1996) is very good
\footnote{
the definition of ``\Lya\ line not associated with metal
systems'' is dependent on the data quality and is therefore not 
rigorous.}.

\section{Metal systems}

In general, we consider a group of absorption lines due to the same ion 
as an individual metal system when the components are blended or
not separated by more than a few tens of km s$^{-1}$. 

The strongest systems in the spectrum of J2233-606 were already
detected by OUT99 and by SAV98. We started from their detections and
looked for weaker components and more ions belonging to the same 
systems. 
Besides, we found 4 new \mgd\ doublets, 7 new \cq\ doublets and a
new associated system showing \cq\ and \nc\ absorptions. 

In the following, we list the systems and give a short description of
their main features. The quoted redshift is the one of the strongest
component or the average redshift in the case of complex systems
(i.e. more than three components). 

All the metal systems are resumed in Table~\ref{tab:metsys}.

\subsection{Intervening systems}

\subsubsection{System at $z_{\rm abs} =-0.000002$}
Strong Na$\:${\small I} $\lambda\lambda\, 5891,5897$ is seen from the
interstellar medium of the Galaxy.

\subsubsection{System at $z_{\rm abs} = 0.0002$}
The complex Ca$\:${\small II} $\lambda\lambda\,3934,3969$ at this
redshift was already observed by OUT99. We detected two more, weaker 
components. The  $\lambda\, 3969$ line is blended with the \nc\
$\lambda\,1238$ feature at $z_{\rm abs} = 2.206$. 

\subsubsection{System at $z_{\rm abs} = 0.41426$}
A single \mgd\ $\lambda\lambda\, 2796,2803$ doublet occurs at this
redshift (OUT99).
The fit with one component yields a relatively poor result, since the
equivalent width ratio of the first line, \mgd\ $\lambda\,2796$,
relative to the \mgd\ $\lambda\, 2803$ line is lower than expected. 
A fit with a double structure does not give a better result. 

Tresse et al. (1999) observed three galaxies at $z \sim
0.4147$ in the 1' field around the quasar. They claim that another object,
closer to the quasar and belonging to the same over-density of galaxies,
could be the responsible for this absorption system.

\subsubsection{System at $z_{\rm abs} = 0.57017$}
Tresse et al. (1999) found at a redshift $z=0.57$ a late-type
spiral galaxy at an impact parameter of $\sim 18\ h^{-1}$
kpc\footnote{ We consider $h = H_0 / 100$ km/s/Mpc and $q_0=0.5$}. 
\huno\ absorption likely associated with this galaxy is seen in the
Lyman series, the fit of the Lyman limit in the STIS spectrum gives
$N($\huno$)\sim 10^{16.8}$ \cm.  

The UVES spectrum shows the associated \mgd\ doublet, with the strongest
component at $z_{\rm abs} = 0.57017$. We fitted the system with 
three components, spanning $\sim 86$ \kms.  

\subsubsection{System at $z_{\rm abs} = 0.58008$}
This new, two component \mgd\ doublet, seen in the UVES spectrum, falls at
the redshift of another galaxy found in the field by Tresse et al. (1999) at an
impact parameter $\sim 112\ h^{-1}$ kpc from the quasar line of
sight. 

\subsubsection{System at $z_{\rm abs} = 0.64498$}
A galaxy at a redshift $z = 0.6465$ was observed at an impact
parameter $\sim 220\ h^{-1}$ kpc from the quasar line of sight by
Tresse et al. (1999). 
The UVES spectrum shows a single \mgd\ doublet at this slightly
lower absorption redshift. 

\subsubsection{System at $z_{\rm abs} = 0.75313$}
A new \mgd\ doublet was found at this redshift in the UVES spectrum. 

\subsubsection{System at $z_{\rm abs} = 1.09169$}
New \cq\ $\lambda\lambda\, 1548,1550$ system, fitted with two
components.

\subsubsection{System at $z_{\rm abs} = 1.32153$}
A new \cq\ doublet in the \Lya\ forest is observed at this
redshift. There is one, clear, unblended component in both $\lambda\, 
1548$ and $\lambda\, 1550$ lines.  On the basis of the absence of the
corresponding Lyman $\beta$ line, we assumed that the whole feature is
\cq\ $\lambda\, 1548$ and not \huno\ \Lya\ and we fitted it with three
components. We tentatively identified a feature showing the same
velocity profile as \cq\ to be the \siq\ $\lambda\, 1393$ absorption, 
while the corresponding \siq\ $\lambda\, 1402$ line is blended with a
strong \huno\ Lya\ complex.  
At a slightly higher redshift, a weak \mgd\ doublet has
been observed showing two stronger components and a flat one in \mgd\
$\lambda\,2796$.  At this redshift, also the lines of \fed\
$\lambda\,2600$, \fed\ $\lambda\,2382$, \cdue\ $\lambda\,1334$,
\sidue\ $\lambda\,1526$ have been detected.

\subsubsection{System at $z_{\rm abs} = 1.3368$}
Tresse et al. (1999) found another quasar in the field at a redshift $z \simeq
1.3360$ and at an angular separation of 44.5'' from the current object.
They measured the \huno\ column density from the \Lya\ and Ly$\beta$
absorption lines, observed in the high resolution portion of the STIS
spectrum. A one component fit gives $N($\huno$)\sim 5\ 10^{15}$
\cm\ and a Doppler parameter $b\sim 50$ \kms.

The UVES spectrum reveals a complex \cq\ doublet absorption at a slightly
higher redshift, covering almost 200 \kms. Even though the system 
falls in the \Lya\ forest, the identification is quite reliable due to
the characteristic profile. 
The \cq\ $\lambda\, 1548$ feature is blended with the possible \sidue\
$\lambda\, 1260$ of the system at $z_{\rm abs} = 1.87$; while the
reddest component of \cq\ $\lambda\,1550$ could be blended with the 
possible \nc\ $\lambda\,1238$ at $z_{\rm abs} = 1.926$. 
We detected also a weak, less extended \mgd\ doublet on the lower
redshift side of the \cq\ absorption and a weak, possible \siq\
doublet corresponding to the highest redshift component of \cq. 
Finally, a fit of the \huno\ \Lya\ line in the STIS spectrum with
the same velocity structure as the \cq\ system gives a total column 
density of $N($\huno$)\sim 3.7\ 10^{16}$ \cm. 

\subsubsection{System at $z_{\rm abs} = 1.4831$}
A complex of five \cq\ $\lambda\lambda\, 1548, 1550$
doublets, spanning 75 \kms, is seen within the \Lya\
forest.  
The identification (by OUT99) is firm thanks to the distinctive 
pattern of the absorption.
No other associated metal line is identified. 

\subsubsection{System at $z_{\rm abs} = 1.5034$}
We cannot confirm the existence of this system tentatively identified
by SAV98. 

The $3\,\sigma$ column density limit on the \mgd\ $\lambda\, 2796$
absorption is $N($\mgd$) \lsim 1.6\times 10^{11}$ \cm .
Other strong low ionization lines like Mg$\:${\small I} 
$\lambda\, 2852$ and \fed\ $\lambda\, 2382$ are not observed and there 
is not the expected strong \Lya\ line in the STIS spectrum. 

\subsubsection{System at $z_{\rm abs} = 1.55635$}
A new, possible, weak \cq\ doublet has been found at this
redshift. The corresponding \Lya\ line has an unexpectedly low column
density, $N($\huno$)\sim 8\times 10^{13}$ \cm. 
 
\subsubsection{System at $z_{\rm abs} = 1.59055$}
This two component \cq\ doublet, separated by 9 \kms, was already
observed by OUT99.

No \siq\ absorption was detected for this system;  
$N($\siq$) \lsim 4.3 \times 10^{11}$ \cm\ is a $3\, \sigma$ 
upper limit on the column density.
The UVES spectrum shows a possible \nc\ doublet feature at the
redshift of the stronger component.   
We fitted the corresponding \Lya\ line with the same components 
as the \cq\ doublet. 

\subsubsection{System at $z_{\rm abs} = 1.73165$}
A possible, weak \cq\ doublet has been found in the UVES spectrum, 
coinciding with a $N($\huno$) \simeq 1.2\times 10^{14}$ \cm\ \Lya\
line. 
 
\subsubsection{System at $z_{\rm abs} = 1.78618$}
This system was already reported by Sealey et al. (1998). 
\cq\ has been observed by SAV98 and by OUT99.

The UVES spectrum does not confirm the presence of Mg$\:${\small I}
$\lambda\, 2852$ tentatively detected by SAV98. 
The $3 \sigma$ limit on this ion is $N($Mg$\:${\small I}$) \lsim 7.2 
\times 10^{10}$ \cm. 
On the other hand, the \siq\ $\lambda\lambda\, 1393,1402$ is
visible in the \Lya\ forest, although there is clear blending with
\huno\ lines. 
The corresponding \huno\ \Lya\ line was fitted with two components for
a total column density, $N($\huno$) \simeq 3.5\times 10^{15}$ \cm. 

\subsubsection{System at $z_{\rm abs} = 1.81565$}
At this redshift, a possible, extremely weak \cq\ doublet absorption
is observed, the $\lambda\, 1550$ line is barely visible. 
We fitted the feature with two components, together with the
corresponding \huno\ \Lya\ line whose total column density is   
$N($\huno$) \simeq 1.4\times 10^{15}$ \cm.

\subsubsection{System at $z_{\rm abs} = 1.8693$}
This complex heavy element system was already reported by OUT99 and by
SAV98. 

The UVES spectrum, besides showing the \siq\ $\lambda\lambda\, 1393, 1402$
doublet, partially resolves the \cq\ $\lambda\lambda\, 1548,1550$
features and shows the associated absorption by \sitre\
$\lambda\,1206$. 
\nc\ $\lambda\,1238$ is blended with a \huno\ \Lya, the $3 \sigma$
limit on \nc\ from the $\lambda\,1242$ line is: $N \lsim 3.4 
\times 10^{12}$ \cm. 
As for the low ionization elements, the \mgd\ doublet is
detected in correspondence of the high redshift components of \siq\ 
and \cdue\ $\lambda\,1334$, \sidue\ $\lambda\lambda\,1190,1193$, 
$\lambda\,1526$, $\lambda\,1260$, Al$\:${\small II} $\lambda\,1670$ 
absorption are also present. 
$N($\fed$) \lsim 4.3 \times 10^{11}$ \cm\ is a $3\,\sigma$ limit on the column 
density of \fed\ $\lambda\,2382$. 
The \huno\ Lya\ line falls in the wavelength range of the UVES
spectrum, the feature is heavily saturated. We obtained an acceptable 
fit with three components, even if the large values of the Doppler
parameters and the velocity profiles of the metal absorptions
suggest a more complex structure. 
We defer the analysis and discussion of this interesting system to a 
following paper (D'Odorico and Petitjean, 2000). 
  
The Mg$\:${\small I} $\lambda\, 2853$ absorption tentatively detected
by SAV98, has been identified with a telluric line by comparing the
spectrum with that of a standard star obtained at comparable
resolution. 

\subsubsection{System at $z_{\rm abs} = 1.92599$}
A prominent \Lya\ was observed at this redshift in the UCLES
spectrum (\cite{outram99}). 

In the UVES spectrum the \cq\ $\lambda\lambda\, 1548, 1550$ doublet is
resolved. We fitted together \cq, \siq\
$\lambda\lambda\, 1393, 1402$, and  \sitre\ $\lambda\, 1206$, with
three components, spanning $\sim 60$ \kms. The \nc\ $\lambda\, 1238$ is
blended with the \cq\ $\lambda\, 1550$ of the system at $z_{\rm abs} =
1.337$.
The possible weaker component is visible, while the existence of the
stronger one is ruled out by the absence of the corresponding \nc\
$\lambda\, 1242$ line. As for the low ionization lines, 
they show a shift of -4 km s$^{-1}$
with respect to the strongest component of the high ionization
absorptions. 
\cdue\ $\lambda\, 1334$ has been fitted with two components, while for
\sidue\ $\lambda\, 1260$ and a possible \mgd\ doublet, 
we fitted only the strongest one. 
The $3\,\sigma$ limit on the abundance of \fed\ based on the
absorption at $\lambda\,2382$ is: $N($\fed$) \lsim 4.9 \times 10^{11}$
\cm. 

A simultaneous fit of \huno\ \Lya, \Lyb, Ly$\epsilon$ and Ly8 has been
carried out.

\subsubsection{System at $z_{\rm abs} = 1.9422$}
This system presents many ions of different elements at different
ionization stages. OUT99 and SAV98 already detected the absorption due
to \cdue\ $\lambda\, 1334$, \sidue\ $\lambda\lambda\, 1260, 1304$,
\sitre\ $\lambda\, 1206$, and \siq\ $\lambda\lambda\, 1393,
1402$, \cq\ $\lambda\lambda\,1548, 1550$, \mgd\ 
$\lambda\lambda\, 2796, 2803  $, \altre\ $\lambda\lambda\, 1854,
1862$, \cuno\ $\lambda\, 1560$, respectively. 

The higher resolution and signal-to-noise ratio of the UVES spectrum
makes it possible to observe three new shallow components in \cq\ that
are also present in \huno\ \Lya.  As for the strongest feature, \cq\
$\lambda\lambda\,1548, 1550$ and \sitre\ $\lambda\, 1206$ are
saturated and have been fitted following the velocity structure of 
\siq\ $\lambda\lambda\, 1393,1402$. 
The $3\,\sigma$ limit on the \nc\ column density is
$N($\nc$) \lsim 1.4 \times 10^{12}$ \cm.  The \huno\ \Lya, Ly$\gamma$,
Ly$\epsilon$ and Ly10 lines have been fitted with the same velocity
structure as the \cq\ doublet for a total column density of
$N($\huno$) \sim 2\times 10^{16}$ \cm\ 
in agreement with the measurement by Prochaska and Burles (1999).

The low ionization lines \cdue\ $\lambda\, 1334$,  
\sidue\ $\lambda\lambda\, 1190,1193$, $\lambda\, 1260$,$\lambda\, 1304$, 
$\lambda\,1526$, \mgd\ $\lambda\lambda\, 2803$, Al$\:${\small II}
$\lambda\, 1670$, and \altre\ $\lambda\, 1854$, have been detected and
fitted with the same velocity profile. 
\mgd\ $\lambda\lambda\, 2796$
was not considered in the fit because its strongest component is
affected by the presence of a telluric absorption line. 
We do not confirm the existence of the \cuno\ $\lambda\lambda\, 1560$ 
absorption tentatively identified by SAV98.
$N($\fed$) \lsim 3.8 \times 10^{11}$ \cm\ is the $3\,\sigma$ limit on
the abundance of \fed\ based on the absorption at $\lambda\,2382$.

A more complete discussion on these latter two interesting metal systems
will appear in a further paper (D'Odorico and Petitjean, 2000).
 
\subsubsection{System at $z_{\rm abs} =2.07728$}
This weak \cq\ doublet was identified by SAV98. 

We tentatively identified \sitre\ $\lambda\,1206$ and \fed\
$\lambda\,2382$ and a more reliable \sidue\ $\lambda\,1260$
absorption.
The $3 \sigma$ limit on the \siq\ $\lambda\,1393$ is: $N($\siq$) \lsim
3.2 \times 10^{11}$ \cm\ 
and that on \cdue\ $\lambda\,1334$, $N($\cdue$) \lsim 1.3 \times
10^{12}$ \cm. 
Unfortunately, the possible \mgd\ $\lambda\lambda\, 2796,2803$
absorptions at this redshift fall in the gap between the two UVES red
arm CCDs.   

\subsubsection{System at $z_{\rm abs} =2.11287$}
New, weak \cq\ doublet fitted with two components. The corresponding
\huno\ \Lya\ line shows a column density $N($\huno$) \simeq 8.1 \times
10^{14}$ \cm. 
  
\subsection{Associated systems}

The associated systems observed in the redshift range $2.198-2.206$ 
have been investigated by Petitjean and Srianand (1999). 
They found that most of the lines in these systems show the signature 
of partial coverage and the covering factor varies from species to
species.  
In general, absolute abundances are close to solar, with the exception
of the [N/C] abundance ratio which is larger than solar. 
This result confirms the physical association of the absorbing gas
with the AGN.

\subsubsection{System at $z_{\rm abs} = 2.19819$}
This associated system shows a strong \cq\ doublet at about $-2000$
\kms\ from the emission peak. 
OUT99 detected and fitted the \nc\ doublet and the \huno\ \Lya\
absorption line.

In the UVES spectrum we found a third, shallow component on the lower
redshift side. \cq\ $\lambda\lambda\, 1548,1550$, \nc\
$\lambda\lambda\,1238,1242$, \osei\ $\lambda\lambda\,1031,1037$,
\huno\ Lya\ and \huno\ \Lyb\ have been fitted together.  The result of
the fit is quite unsatisfactory ($\chi^2 = 9.5$), due to the unusual
ratio between the lines of the doublets. In particular, the \osei\
lines show a flat bottom - as if they were saturated - but a non zero
residual flux, a clear signature of partial coverage.  The $3\,\sigma$
limit on the \siq\ $\lambda\,1393$ is $N($\siq$) \lsim 5.2\times
10^{11}$ \cm. 

\subsubsection{System at $z_{\rm abs} = 2.20106$}
This new, broad and shallow system shows the \nc\
$\lambda\lambda\,1238,1242$ and \cq\ $\lambda\,1548$ lines.
\cq\ $\lambda\,1550$ and \osei\ $\lambda\lambda\,1031,1037$ are blended
in the forest. 
The $3\,\sigma$ limit on \huno, from the absence of detectable \huno\
\Lya, is $N($\huno$) \lsim 3.6 \times 10^{11}$ \cm.   

\subsubsection{System at $z_{\rm abs} = 2.2064$}
A multi-component, broad \cq\ absorption in the redshift range between
about $-1400$ \kms \ and $-1100$ \kms\ from the emission peak was
observed by SAV98. OUT99 fitted the \nc\ $\lambda\lambda\,
1238,1242$ and the \huno\ \Lya\ with three components, the fit is
complicated  by uncertainty in the precise shape of the \Lya\ \& \nc\
emission line continuum, also, the \nc\ $\lambda\, 1242$ from the
$z_{\rm abs} = 2.198$ system lay on top of the \nc\ $\lambda\, 1238$. 

We performed a simultaneous fit of the \cq, \nc, \osei\ doublets and 
the \huno\ \Lya\ with five components. 
The $3\,\sigma$ limit on the \siq\ column density is $N($\siq$) \lsim
3.3 \times 10^{11}$ 
\cm. 

\acknowledgments
We are greatly indebted to all the people involved in the conception,
construction and commissioning of the UVES instrument, without whom
this project would have been impossible. 
We are grateful to the members of the ESO Data Management Division
Pascal Ballester, Sebastian Wolff and Andrea Modigliani for the 
development of the
MIDAS-based UVES pipeline which was used to reduce these data.
We thank J. Bergeron, S. D'Odorico, T.-S. Kim and S. Savaglio for
enlightening discussions.
This work has been conducted with partial support by the TMR programme
Formation and Evolution of Galaxies set up by the European Community
under the contract FMRX-CT96-0086.
\clearpage
\begin{deluxetable}{lccc}
\tablewidth{400pt}
\tablecaption{Journal of the observations of J2233-606}
\tablehead{
\colhead{Date} &
\colhead{Wavelength Range} &
\colhead{Dichroic} &
\colhead{Exposure Time} \\
\colhead{} &
\colhead{\AA} &
\colhead{} &
\colhead{s} 
}
\startdata
1999 Oct 08 & 3050-3875  & DIC1 & 3600 \nl
1999 Oct 08 & 3050-3875  & DIC1 & 3600 \nl
1999 Oct 11 & 3050-3875  & DIC1 & 3000 \nl
1999 Oct 12 & 3050-3875  & DIC1 & 3000 \nl
1999 Oct 12 & 3050-3875  & DIC1 & 3000 \nl
1999 Oct 10 & 3760-5004  & DIC2 & 2700 \nl
1999 Oct 10 & 3760-5004  & DIC2 & 2700 \nl
1999 Oct 10 & 3760-5004  & DIC2 & 3300 \nl
1999 Oct 16 & 3760-5004  & DIC2 & 3600 \nl
1999 Oct 08 & 4795-6815  & DIC1 & 3600 \nl
1999 Oct 08 & 4795-6815  & DIC1 & 3600 \nl
1999 Oct 12 & 4795-6815  & DIC1 & 3000 \nl
1999 Oct 12 & 4795-6815  & DIC1 & 3000 \nl
1999 Oct 12 & 4795-6815  & DIC1 & 3000 \nl
1999 Oct 10 & 6725-10000 & DIC2 & 2700 \nl
1999 Oct 10 & 6725-10000 & DIC2 & 2700 \nl
1999 Oct 10 & 6725-10000 & DIC2 & 3300 \nl
1999 Oct 16 & 6725-10000 & DIC2 & 3600 \nl
\label{tab:obs}
\enddata
\end{deluxetable}
\clearpage
\begin{deluxetable}{lccccc}
\tablewidth{400pt}
\tablecaption{List of metal absorption systems}
\tablehead{
\colhead{redshift\tablenotemark{a}} &
\colhead{$\log N($\huno$)$\tablenotemark{b}} &
\colhead{$\log N($\siq$)$\tablenotemark{b}} &
\colhead{$\log N($\cq$)$\tablenotemark{b}}  &
\colhead{$\log N($\mgd$)$\tablenotemark{b}} &
\colhead{\#\tablenotemark{c}}
}
\startdata
-0.000002\tablenotemark{d} & \nodata & \nodata & \nodata &\nodata & 1 \\ 
0.0002\tablenotemark{e} & \nodata & \nodata & \nodata & \nodata & 4 \\ 
0.41426 & \nodata & \nodata & \nodata & 12.79   & 1 \\
0.57017 & \nodata & \nodata & \nodata & 12.55   & 3 \\ 
0.58008 & \nodata & \nodata & \nodata & 12.60   & 2 \\ 
0.64498 & \nodata & \nodata & \nodata & 12.34   & 1 \\ 
0.75313 & \nodata & \nodata & \nodata & 11.27   & 1 \\
1.09169 & \nodata & \nodata & 13.53   & $<$ 11.3 & 1 \\
\enddata
\tablenotetext{a}{See definition in \S 5} 
\tablenotetext{b}{Logarithmic total column density}
\tablenotetext{c}{Number of components fitted to the \cq\ absorption
feature or to the \mgd\ absorption feature, if \cq\ is not observed}
\tablenotetext{d}{Galactic \nauno\ $\lambda\lambda\, 5891,5897$}
\tablenotetext{e}{Galactic \cadue\ $\lambda\lambda\, 3934,3969$}
\tablenotetext{f}{Associated systems}
\end{deluxetable}
\clearpage
%
%
%

\clearpage

\end{document}